\documentclass[lettersize,journal]{IEEEtran}
\usepackage{amsmath,amsfonts}
\usepackage{algorithmic}
\usepackage{array}
\usepackage[caption=false,font=normalsize,labelfont=sf,textfont=sf]{subfig}
\usepackage{textcomp}
\usepackage{stfloats}
\usepackage{url}
\usepackage{verbatim}
\usepackage{graphicx}
\usepackage{cite}
\usepackage{xspace}
\usepackage{epsfig}
\usepackage{xcolor}
\usepackage{makecell}
\usepackage{amsmath}
\usepackage[ruled,vlined]{algorithm2e}
\newcommand*{\revise}{\textcolor{black}}
\newcommand*{\highlight}{\textcolor{black}}

\begin{document}

\title{Fast-OverlaPIM: A Fast Overlap-driven Mapping Framework for Processing In-Memory Neural Network Acceleration}

\author{Xuan Wang\textsuperscript{\textsection}, Minxuan Zhou\textsuperscript{\textsection}, and Tajana Rosing~\IEEEmembership{Fellow,~IEEE,}
\thanks{X. Wang and M. Zhou contributed equally to this research.}
\thanks{}
\thanks{}
}

\markboth{The IEEE TRANSACTIONS ON COMPUTER-AIDED DESIGN OF INTEGRATED CIRCUITS AND SYSTEMS}%
{Shell \MakeLowercase{\textit{et al.}}: A Sample Article Using IEEEtran.cls for IEEE Journals}

\maketitle

\newcommand{\Design}{Fast-OverlaPIM\xspace}

\newcommand{\squeezeup}{\vspace{-0mm}}
\newcommand{\squeezeupless}{\vspace{-0mm}}

\begin{abstract}
Processing in-memory (PIM) is promising to accelerate neural networks (NNs) because it minimizes data movement and provides large computational parallelism. Similar to machine learning accelerators, application mapping, which determines the operation scheduling and data layout, plays a critical role in the NN acceleration on PIM. 
The mapping optimization of previous NN accelerators focused on optimizing the latency of sequential execution. However, PIM accelerators feature a distinct design space of application mapping from conventional NN accelerators, due to the spatial execution of NN layers across different memory locations. This enables opportunities for overlapping execution of consecutive NN layers to improve the latency, where the succeeding layer can start execution before the preceding layer fully completes the computation.
In this paper, we propose \Design framework that incorporates the computational overlapping optimization into the DNN mapping exploration process on PIM architectures. \Design includes analytical algorithms for fast and accurate overlap analysis. Furthermore, it proposes a novel mapping search strategy and a transformation mechanism to enable efficient design space exploration on the overlap-based mapping for the whole network. Our framework demonstrates a significant improvement in runtime performance from 3.4$\times$ to 323.1$\times$ compared to the previous state-of-the-art overlap-based framework. Our experiments show that \Design can efficiently produce mappings that are 4.6$\times$ to 18.1$\times$ faster than the state-of-the-art mapping optimization framework under the same architecture constraints.
\end{abstract}


\section{Introduction}

\revise{Deep neural networks (DNNs) have grown in popularity in a wide range of areas, such as image and video recognition \cite{touchgo}, \cite{videohuman}, \cite{image1} IoT devices \cite{iot1}, \cite{iot2}, \cite{iot3} and natural language processing \cite{nlp1}, \cite{nlp2}. With continuously increasing network size \cite{resnet} due to the demand for high accuracy, memory, and computation demands are rising dramatically for emerging neural networks (NNs), significantly slowing down the performance.}

\begin{figure}[t!]
    \centering   
    \epsfig{file=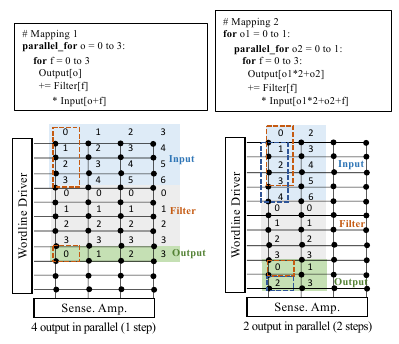, width=0.95\columnwidth}
    \caption{Data layout of mappings of a 2D convolution.}
    \label{fig:pim_mapping}
\end{figure}

\revise{Processing in-memory (PIM) has emerged as an acceleration solution to boost the performance of computational heavy applications, including DNNs.}
PIM exploits the large chunk of memory resources as both storage and processing elements to realize \revise{high compute and memory throughput simultaneously. The large capacity of memory allows the allocation of separate memory resources for different parts of the application, which significantly improves the computational throughput \cite{floatpim} \cite{gram}\cite{transpim}. Therefore, existing large-scale PIM accelerators spatially distribute all NN layers to exploit PIM parallelism \cite{felix} \cite{pimdnn1} \cite{pimdnn2}. The spatial processing in PIM architectures leads to a unique design space for NN acceleration, which determines the execution order and distribution of DNN operations on the hardware \cite{timeloop} \cite{pim-dl} \cite{sparseloop}.
Figure~\ref{fig:pim_mapping} shows an example of two mappings for a 2D convolution on a PIM-enabled memory block. Different mappings vary in latency, data layout, and the order of producing outputs and consuming inputs. In large-scale DNNs, the design space for mapping DNNs onto the hardware accelerator is large. The mapping choices will be influential to latency and memory resource allocation.}

\revise{To tackle the mapping optimization problems, accelerators developed optimization for the mappings based on hardware configurations\cite{eyeriss}\cite{MAESTRO}. However, frameworks for conventional architecture \cite{timeloop} \cite{interstellar} \cite{ruby} \cite{huang2021cosa} \cite{anadnn} assume the accelerator processes one layer at a time, do not apply to spatially distributed acceleration on PIM architecture. PIM solutions are proposed \cite{puma} \cite{prime} \cite{yang_design} \cite{z-pim} \cite{pim-dl} to consider optimization enabled by spatial memory. However, most PIM accelerators use heuristics for mapping and lack support for overlap optimization. In practice, the execution of a layer can start earlier when part of its input has been output from the preceding layer(s). Therefore, we overlap the execution of different layers to improve the performance.} We call this improvement computational overlap optimization between consecutive layers. \revise{The lack of consideration for the computational overlap enabled by spatially distributed DNN acceleration between two consecutive layers leads to the undervaluing of the high parallelism of PIM.} As a result, when considering the overlapping, the existing PIM-based framework, which only considers the end-to-end sequential latency, may generate sub-optimal mappings.

\revise{This work proposes an optimized PIM-based DNN mapping framework, \Design, to enable efficient overlap-based mapping optimization for the
whole network on PIM architectures. Overlap analysis necessitates the generation of fine-grained data spaces, where a data space represents the range of data processed in a memory partition \revise{(e.g., bank, block)} at each time step, to determine the ready time step for each data space.} Considering each memory contains hundreds of banks and each DNN layer requires thousands of time steps, generating fine-grained data space information can be prohibitively costly. Moreover, the naive approach of analyzing output/input dependency by exhaustively comparing all data spaces among two consecutive layers in OverlaPIM also becomes the bottleneck. \revise{Meanwhile, mapping search becomes slow with the integration of overlap analysis so the exhaustive search strategy could be impossible. In response to these challenges, we propose several novel strategies them \Design.}
\begin{itemize}
\item We modify the algorithm in OverlaPIM to a lightweight algorithm for generating fine-grained data spaces. 
\item \revise{We propose an analytical algorithm to determine the overlapping performance without exhaustive comparison.} 
\item  We propose a transformation mechanism that transforms an analyzed mapping into overlap-friendly mappings with a trivial overhead. We effectively increase the search space of the framework in a similar amount of time. 
\item We propose a novel search strategy that provides additional insights for map space exploration. Rather than initiating the search for map spaces from the first layer in the DNN, we opt to use the last layer or middle layers as the starting point for our whole network optimization. 
\item We implement the proposed framework in an open-source DNN mapping framework, Timeloop \cite{timeloop}, and compare \revise{it} against state-of-the-art mapping optimization without consideration of overlap. The analytical algorithms provided 3.4$\times$ to 323.1$\times$ runtime performance improvement as compared to exhaustive comparison for data dependencies analysis in OverlaPIM\cite{overlapim}. Our evaluation of popular DNNs indicates  \Design can produce mappings that are 4.6$\times$ to 18.1$\times$ faster than the mappings optimized by existing methods~\cite{timeloop,pim-dl}.
\end{itemize}

\section{Related Work}
\subsubsection{ASIC solutions}
\revise{Chen et al. \cite{eyeriss} reduce DNN latency by proposing the row-stationary mapping of Eyeriss to minimize data movement with weight sharing and data fusion. Parashar et al \cite{timeloop} explore a large design space and develop the Timeloop framework to accommodate flexible architectures while searching for the optimal mapping.} Timeloop employs a near-exhaustive search strategy to identify efficient mappings, taking into account the size of operation spaces to calculate mapping latencies. Other works \cite{interstellar}\cite{ruby} \cite{MAESTRO} also propose ASIC frameworks for mapping optimization through searching. Huang et al. \cite{huang2021cosa} and Zhao et al. \cite{anadnn} improve machine learning latency by solving optimization problems. Although the above approaches increase performance, their emphasis lies on single-layer mapping optimization without considering data movement between consecutive layers. Consequently, their designs may not fully support the spatial architecture of PIM.

\subsubsection{In-Memory solutions}
Ankit et al. \cite{puma} propose Puma, a data-flow-based accelerator that allocates MVM (Matrix-Vector Multiplication) operations in DNNs on a spatial PIM architecture through compiler optimizations. Zhou et al. \cite{pim-dl} improve DNN performance through the development of PIM-DL. \revise{This approach integrates global optimization with per-layer mapping data layout optimization at the compiler level to minimize data movements between consecutive DNN layers.} However, the existing PIM frameworks are grounded in the assumption that only a single layer is processed at a time. \revise{They overlook the potential cross-layer optimization for the computational overlap facilitated by spatially distributed PIM computational resources for DNN acceleration, which results in the undervalue of PIM's high parallelism.}

\subsubsection{OverlaPIM framework}
Our previous work, OverlaPIM \cite{overlapim}, to the best of our knowledge, is the first to consider computational overlapping during mapping optimization for NN accelerations in PIM architecture and were 2.1$\times$ to 4.1$\times$ faster than Timeloop \cite{timeloop} mapping optimization framework on workload ResNet-18 \cite{resnet} and VGG-16 \cite{vgg}. OverlaPIM decomposes inputs, outputs, and weights into smaller \textit{data spaces} to be processed in each memory element at specific time steps. Nevertheless, OverlaPIM relies on exhaustively comparing all data spaces between two consecutive layers for overlap-based performance analysis to estimate the overlapped performance. Given the substantial size of data spaces resulting from DNN mappings, these analysis processes can become prohibitively expensive and create a large runtime bottleneck. As a result, the exploration scope for each layer was constrained, leading to the significant underperformance of the whole network optimization in the OverlaPIM framework.

\section{Background and Motivation}
\subsection{Preliminaries for Processing In-memory Architecture} 
\label{sec:pim_background1}
\revise{PIM} has been widely used for memory-intensive and compute-intensive applications because of its extensive bandwidth for computing and ability to minimize data movement. \revise{Various PIM-based architectures have been introduced to support parallel operations inside memory with different memory technologies, such as SRAM\cite{wang2019computesram}, DRAM\cite{drisa,simdram, gao2019computedram,ali2019memory}, ReRAM\cite{dpsim,floatpim,reram1}, spintronics-based magnetic memories \cite{spin-electronics}. This work focuses on the digital PIM paradigm, which stores the conventional digital representation (i.e., one binary bit per cell) in memory. A common underlying foundation for digital PIM is row-parallel bit-serial operations \cite{ali2019memory}, which make use of the common bit-line circuits shared in memory blocks to process all columns in selected operand rows in a single step. Supported with universal bit-wise operations (e.g. AND, NOR, majority operation, etc.), PIM memory enables general computations for DNN \cite{ambit,ambit2,simdram}. Our baseline system is the bit-serial row-parallel PIM architecture~\cite{gao2019computedram} based on HBM2 DRAM~\cite{hbm}, which supports universal bit-wise operations in the DRAM banks with activate-activate-precharge~\cite{simdram} and performs majority-based addition (MAJ)~\cite{ali2019memory}.}

\highlight{
In addition to the baseline DRAM system, \Design can be aligned with other digital PIM-based architectures through minimal modifications for the performance model and customized architecture configurations as inputs. In Section~\ref{sec:config} and Section~\ref{sec:perf_model}, we will furtherdiscuss the possibility of extending \Design to support analog PIM, which represents fixed operands (i.e., neural network weights) in memory cells and process computation by applying analog signals (i.e., current) as other operands and sensing the result signals~\cite{analog,mnsim}.}

\begin{figure}[t!]
    \centering
    \epsfig{file=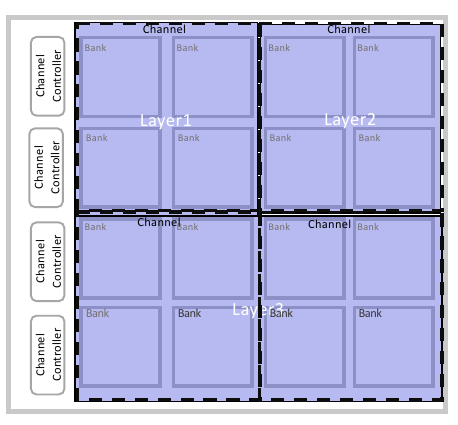, width=0.7\columnwidth}
    \caption{\revise{High-level overview example of DRAM-based PIM architecture.}}
    \squeezeup
    \label{fig:pim_structure}
\end{figure}

\subsection{PIM Acceleration for DNN} \label{sec:pim_background2} 
\revise{Floating-point matrix operations, such as additions and multiplications,  are the bottleneck for DNNs. For each layer, weights and inputs are loaded into memory spaces for computation. After execution,} outputs are reallocated to the input memory space for the next layer without transferring between memory and processing units. \revise{As shown in Figure~\ref{fig:pim_mapping},} we allocate memory rows to different bits of operand and result vectors to support an in-memory computation for each addition and multiplication operation. The example only shows single-bit values for input, filter, and output. In practice, each data may take multiple rows. Once the operand and result vectors are aligned, the memory issues a sequence of universal bit-wise operations to generate the result vector. Such bit-serial operations achieve extensive parallelism because we can simultaneously process all columns in memory rows in different memory blocks (e.g. Bank and Channel).

\highlight{
Figure~\ref{fig:pim_structure} shows an example of high-level PIM architecture for DNN acceleration. The memory is organized into multiple channels, each containing several banks. Each bank is a 2D memory array, which is the basic unit of PIM computing. We note that this spatial layout is general to support various technologies, such as digital bit-serial computing or analog computing.}
\revise{To accelerate NN, each layer may be pre-allocated with several memory banks spatially to exploit parallelism.} For networks with larger depths, fully completed layers will be replaced by new subsequent layers. \revise{Thus, through pre-allocating memory resources for different layers given memory constraints, the PIM-based accelerators can accommodate the whole DNN network, different from ASIC-based accelerators that usually process one layer at a time~\cite{timeloop}.} The whole network execution avoids costly off-chip data communication. 

\revise{Achieving optimized performance PIM acceleration on NN requires sophisticated design choices on software-to-hardware mapping, which determines the operation scheduling and data allocation of a specific DNN across the hardware resources~\cite{timeloop}. Each mapping has a direct relation to a specific data allocation and operation schedule on the given hardware, which has hierarchical storage. The per-layer optimization (i.e. process one layer at a time) for customized accelerators may lead to sub-optimal performance for PIM architectures with sufficient resources to process many layers simultaneously.}

\subsection{Effect of Computational Overlap}
\label{sec:motivation_s}
 \revise{Figure~\ref{fig:pim_mapping} shows an example of two different mappings in PIM architecture.} The mappings are described using the syntax of Timeloop~\cite{timeloop}. \revise{Mapping1} parallelizes computations of all 4 outputs by exploiting memory resources and spreading data in different memory columns for each layer; \revise{Mapping2} parallelizes two output tiles where each tile sequentially processes 2 outputs by dividing memory into two consecutive NN layers. \revise{Mapping1 and Mapping2 have different data layouts, which affects both the latency for processing the layer and the order of producing outputs and consuming inputs through different allocations.} Given the layer-wise data dependencies, the order of generating outputs and utilizing inputs significantly affects the ability for computational overlap optimization.

\begin{figure}[t!]
    \centering   \epsfig{file=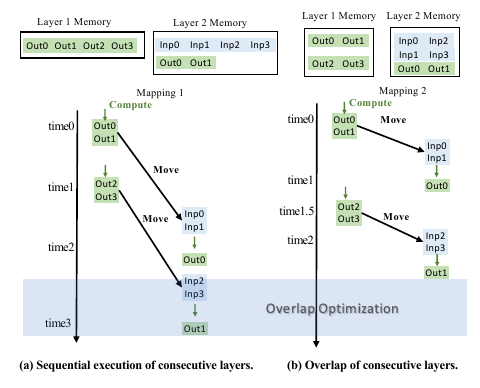, width=\columnwidth}
    \caption{Comparison between conventional overlap-friendly layer execution.}
    \label{fig:overlap_comp}
\end{figure}

\revise{Figure}~\ref{fig:overlap_comp}(a) and \revise{Figure}~\ref{fig:overlap_comp}(b) demonstrate overlap optimization through the comparison of execution times and memory allocation between sequential execution and paralleled spatially execution based on the different mapping choices of two consecutive layers. We divide the computation of each layer into multiple time steps, where each time step processes several operation spaces in parallel (spatially processed in different memory blocks). In \revise{Figure}~\ref{fig:overlap_comp}(a), larger PIM-based memory is allocated for each layer so that Layer1 achieves full completion at time step1. However, Layer2 needs to wait until after time step1 to start execution and the whole processing for the two layers will be finished at step3. But in \revise{Figure}~\ref{fig:overlap_comp}(b), smaller memory is allocated for each individual in order to trade for parallelism between Layer1 and Layer2 so that Layer1 can only be fully completed at time step1.5 due to limited memory allocation and computation resources. However, Layer2 can overlap the computation of its output0 with Layer1 at time step , where Laye 1 completes the computation of operation spaces output0 and output1 (input0 and input1 in Layer2). Output1 of Layer2 can also be overlapped. Layer2 can finish execution before time step3 and gain the improvement from overlap optimization. Such overlap optimization can bring better performance than the performance considered by the existing framework considering the increasing per-layer size of modern DNNs.

\subsection{Overlap-based Optimization for PIM DNN Acceleration}

\begin{figure}[t!]
    \centering
    \epsfig{file=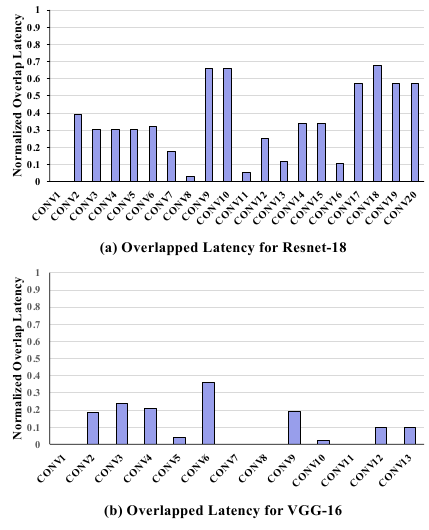, width=\columnwidth}
    \caption{The normalized overlapped latency for all layers in ResNet-18 and VGG-16 optimized by existing framework~\cite{timeloop} - higher means better overall performance.}
    \squeezeup
    \label{fig:motivation}
\end{figure}

To harness the overlap optimization introduced in~\ref{sec:motivation_s}, we initially explored the potential for overlapping in mappings generated with conventional DNN frameworks. \revise{In Figure~\ref{fig:motivation}, we display the experimental result of PIM acceleration for ResNet-18 and VGG-16 where we use the state-of-the-art open source frameworks, Timeloop~\cite{timeloop}, to search for the best mapping (mapping with the lowest latency) layer by layer.} \revise{We added the analysis of the computation overlapping of consecutive layers on the top of Timeloop and reduced the overlapped computations from the original latency if and only if the input for all operation spaces of the following layer becomes ready in a specific time step.} Therefore, not all overlapped data spaces lead to a latency reduction. This figure is produced by calculating the normalized latency of the overlapped computation in the experiment, where a higher value indicates better performance with more overlapped computations. \revise{If we naively search for the best non-overlap performance mapping for each layer, the overlapped latency varies significantly from layer to layer for both workloads.} For ResNet-18, 10 out of 20 layers only have very limited overlapping ($\leq30\%$) while other layers can overlap a significant portion of their computations (31\% to 68\%). And for VGG-16, 5 out of 13 layers cannot gain latency improvement at all and 4 out of 13 layers only have trivial overlapping ($\leq10\%$) while the rest layers have some overlapping (11 \% to 37 \%). \revise{Therefore, there exists great potential to optimize the performance by optimizing the DNN mapping on PIM based on overlaps of computations between consecutive layers.}

\section{Fast-OverlaPIM Framework} \label{sec:design}
In this work, we propose \Design, a PIM mapping framework for DNN with the computational overlapping optimization between consecutive layers, \revise{considering} hardware architecture constraints and software algorithm optimization.

\subsection{The Flow of Fast-OverlaPIM} \label{sec:flow}
\begin{figure}[t!]
    \squeezeup
    \centering
    \epsfig{file=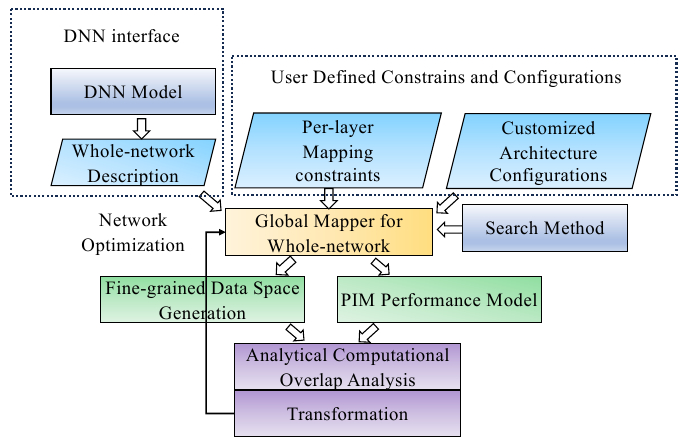, width=1.0\columnwidth}
    \caption{The overview of \Design framework.}
    \squeezeup
    \squeezeup
    \label{fig:overview}
\end{figure}

Figure~\ref{fig:overview} shows the key components of \Design to enable overlap-based optimization. First, we design a new interface to configure the hardware and optimization procedure of the whole network. Second, we add a PIM performance model to support the accurate evaluation of mapping on PIM architectures. Third, we develop a lightweight fine-grained data space generation algorithm. This process generates detailed data spaces over time on various memory components, facilitating efficient preparation for overlap analysis and optimization within a relatively short runtime. Then, we propose a fast analytical computational overall analysis algorithm to resolve the output-input data dependencies between the consecutive layers to determine the overlapping performance. We propose a transformation algorithm that can significantly increase the search capability of whole DNN optimization with trivial overhead. Moreover, we introduce novel search strategy alternatives with the capability of starting from layers other than the numeric first layer to explore larger design spaces and to \revise{improve} the performance of searched mappings.

\subsection{DNN Interface and Configurations}~\label{sec:config}
To develop a user-friendly framework for the whole network mapping, we implemented the DNN interface that takes the whole DNN model as input and gives \revise{the whole-network layer information} as outputs to serve as configurations for \Design. \revise{\Design interface extracts the size parameters of the input, filter, and output for each NN layer. The extraction results are in \Design readable format and processed by \Design in sequential order as a whole-network description. These interfaces will allow the users to configure the whole network for the overlap analysis between consecutive layers as a global mapper without much manual effort.} \revise{We also enable the new interface to take the description of per-layer mapping constraints as inputs to assist with the mapping search. User-defined mapping constraints provide additional information for tiling and allocating matrix workloads onto hardware components.}

\begin{figure}[t!]
    \squeezeup
    \centering
    \epsfig{file=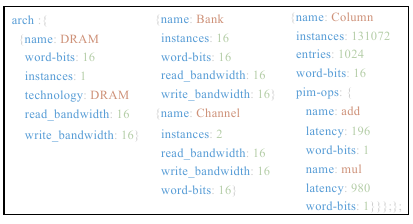, width=0.7\columnwidth}
    \caption{\revise{Sample user-customized architecture configurations of DRAM-based PIM architecture.}}
    \squeezeup
    \label{fig:arch_config}
    \squeezeup
\end{figure}

\highlight{For the hardware configuration, the interface takes the user-customized PIM model as input. For each memory level, we can configure the number of parallel instances and word bits to represent the operation throughput; the read bandwidth and write bandwidth can be specified to model different intra-memory links for data movement.}
\revise{Figure~\ref{fig:arch_config} shows an example of DRAM-based PIM architecture configuration. The configuration is constructed following the hierarchical tree structure of storage elements. The read/write bandwidth of each channel is 16 bytes. If a level does not have read/write bandwidth (e.g., Column in the example), the upper level (i.e., Bank) handles the data movement. We can specify the supported PIM operations at each level to emulate different PIM structures. The example shows the configuration to support bit-serial row-parallel operations, where all column instances can compute 1-bit addition or multiplication simultaneously.}  
\highlight{In Section~\ref{sec:otherpim}, we illustrate how to exploit \Design as a general-purpose PIM-based mapping optimization framework that can support various PIM-based architectures.}

\squeezeup
\squeezeup
\squeezeup
\squeezeup

\subsection{PIM Performance Model}~\label{sec:perf_model}
We implemented a new PIM performance for latency performance evaluation because the Timeloop~\cite{timeloop} performance model only considers the compute, read, and write latency. These statistics are insufficient for PIM performance evaluation which requires data movement. In the context of the PIM architecture the PIM architecture for \Design, we focus on spatially distributed PIM acceleration for our network optimization based on bit-serial row-parallel processing~\cite{gao2019computedram}, as introduced in Section~\ref{sec:pim_background1} and Section~\ref{sec:pim_background2}. We allocate a fixed amount of memory for each individual DNN layer, and the framework determines the data layout based on the mapping. We allocated a fixed amount of memory for each DNN layer, and the framework optimizes the data layout. Based on the mapping, we place \revise{the} filter of all layers and input data for the first layer. After the completion of the execution for each layer, we move its output to the corresponding memory locations of the input for the next layer. Our PIM performance model estimates the performance of such processing flow. 

\revise{To represent the computation latency of PIM, we replace the read/write operations in \cite{timeloop} with the data movements required by PIM executions.} Data movements include the output-input inter-layer data transfer and the data movements for reductions of partial sums located in different columns. \revise{Specifically, for each MAC operation in a memory bank, we model the operation using three steps: 1) the element-wise multiplications for partial products, 2) memory read/write for transposition, and 3) serial additions for reduction. The latency and energy of read/write are similar to the normal DRAM operation, with activation, precharging, and read/write DRAM commands. Each full addition requires $4n+1$ activate-activate-precharge (AAP) operations, where n is the number of bits per value (16-bit in our experiments). Each multiplication consists of sequential full additions.}

\squeezeup
\squeezeup
\begin{figure}[t!]
    \squeezeup
    \centering
    \epsfig{file=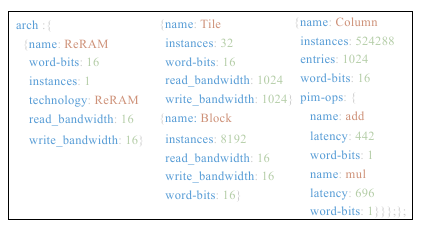, width=0.8\columnwidth}
    \caption{\highlight{Sample user-customized architecture configurations of ReRAM-based PIM architecture \cite{floatpim}.}}
    \squeezeup
    \label{fig:reram}
    \squeezeup
\end{figure}
\squeezeup
\squeezeup

\highlight{\subsection{Applicability to General PIM Architectures}}~\label{sec:otherpim}

\highlight{Our overlap-based analysis and optimization are applicable to other spatial DNN accelerations, including PIM architectures with different technologies or ASIC-based spatial accelerators. For instance, Figure~\ref{fig:reram} illustrates the support for FloatPIM, a ReRAM-based digital PIM architecture that exploits bit-serial row-parallel computations for DNN training and inference\cite{floatpim}. FloatPIM is compatible with any bipolar resistive technology, which is the most commonly used in existing non-volatile memories (NVMs). By adjusting the memory hierarchy level and the corresponding number of instances in the user-defined configuration, the FloatPIM architecture can be represented in \Design. Adjusting the add and multiply latency ensures the corresponding cycle-latency measurement for the ReRAM-based PIM. Furthermore, we can adjust the read and write bandwidth in the memory level to represent the customized intra-tile links in FloatPIM.
These configuration adjustments are similarly generalizable to other PIM-based architectures.}

\highlight{While the current implementation of \Design focuses on digital-based PIM architectures, which allocate all data in the conventional digital memory, we can extend it to support analog PIM by configuring architectural configuration input and performance model. In analog PIM\cite{analog}\cite{mnsim}, only the filter is allocated directly in memory, and input will be loaded as the analog signal to the memory for computation. However, the output is still generated in the temporally sequential order for each spatially distributed PE. Thus, the partial input for the succeeding layer can still be finished and loaded to the next filter before the proceeding layer is fully completed. Therefore, we can modify the architecture configuration to model the analog PIM memory and implement a new performance model to consider the analog-specific operations (e.g., ADC, DAC, etc.).}

\subsection{Data Space Decomposition}
\label{sec:data}
DNN mappings describe the way in which inputs, outputs, and weight are split into data spaces at each level of the memory \revise{hierarchy} and at each time step. We utilize a consistent representation of data spaces and loop decomposition to facilitate mapping generation, fine-grained data space generation, and computational overlap analysis.

\Design exploits the conventional 7D-loop representation for the DNN layer. We use the convolution layer as the example, where $R$ and $S$ are the height and width of weight, $P$ and $Q$ are the height and weight of output, $C$ is the number of input channels, $K$ is the number of output channels, and the number of inputs or batch size is represented by $N$. With this parameterized representation, we define the output data space as a 4-D tensor $[N, K, P, Q]$ and the input data space as $[N, C, P + R - 1, Q+S-1]$ for further input/output overlapping analysis on different mappings. This can represent the whole DNN model because CONV and FC (Fully-Connected) layers dominate the computations and data movement. Figure~\ref{fig::generation} shows an example of data spaces for the mapping on a two-level memory. For simplicity, we ignore the dimension of $N$ in the later discussion. 

In Figure~\ref{fig::generation} (a), each mapping, with a specific loop decomposition and permutation, can be translated into data spaces that are spatially and temporarily distributed. \revise{Specifically, the spatial distribution (i.e., $parallel\_for$) means data spaces are split and allocated to different hardware instances.} The temporal distribution (i.e., $for$) happens when a data space of a hardware instance is further decomposed into smaller data spaces. These smaller data spaces are sequentially processed by the corresponding hardware component in multiple temporal steps, where the instance processes a temporal data space in each step. \revise{In Figure~\ref{fig::generation},} the whole output data space is spatially decomposed into two channel-level spaces, each of which is further decomposed into two temporal steps. Similarly, each channel-level temporal step is decomposed into 2 bank-level spaces, where each consists of $P*Q$ temporal steps. In total, each bank in this example consists of $2*P*Q$ temporal steps.

\subsection{Fine-grained Data Space Generation} \label{sec:space}

\revise{Analysis of the overlap between layers requires the comparison of the detailed output/input data spaces for each memory hierarchy level across DNN layers.} However, previous works avoid generating fine-grained data spaces due to runtime limitations. Timeloop~\cite{timeloop} only collects a small portion of data spaces for size measurement. Since Timeloop generates data spaces from recursive function calls, collecting all data spaces is unacceptably expensive for both runtime efficiency and memory consumption and makes overlap analysis impossible.

\revise{To reduce the complexity, we propose a lightweight fine-grained generation algorithm to infer all spatial and temporal data spaces through analytical observation and formulation. In Figure ~\ref{fig::generation}(b)(c), Timeloop~\cite{timeloop} generates partial data spaces in the light green and yellow. Our analytical algorithm utilizes these data spaces as reference points to generate fine-grained data spaces. In order words, we identify and generate the $4\times(PQ-2)$ number of previously un-generated data space, as highlighted within the blue rectangle.}

\revise{
\begin{figure}[t!]
    \centering
    \epsfig{file=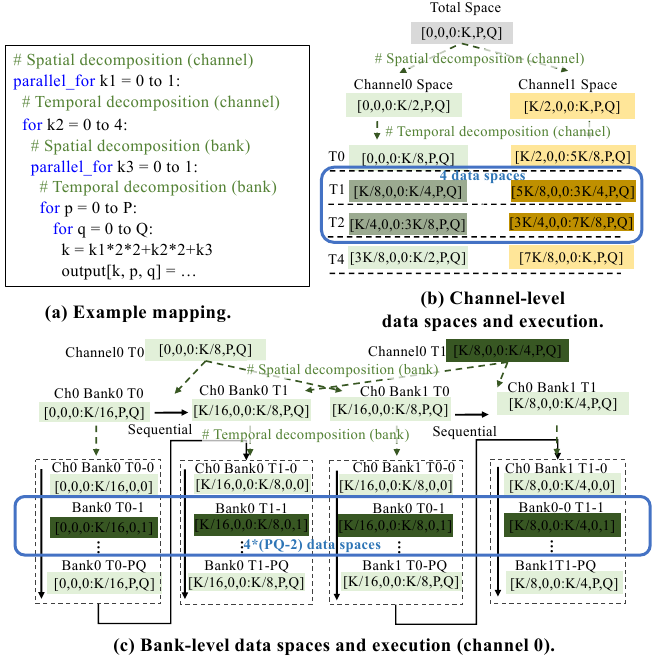, width=1\columnwidth}
    \caption{\revise{Fine-grained Data Space Generation.}}
    \label{fig::generation}
\end{figure}
}

\revise{We propose a lightweight data space analysis based on the observation that the size of data spaces remains the same at each hardware level.} The dimension value of data spaces changes periodically corresponding to the inner loop iteration, where each loop level increments one data space dimension. Since these increments are well-patterned with the fixed total number of data spaces and stable size of each individual data space, we conclude a lemma to define the relationship between the index increment of the specific loop level $n$ and its corresponding time step increment:
\begin{equation} \label{eq:1}
    \ G(n) = \prod_{j = q}^{n-1} num_{j}
\end{equation}
where $num_{j}$ is the number of iterations in the $j^{th}$ temporal loop, and $q$ is the lower bound of loops in the target hardware level where the current loop iteration belongs.

Then, our analysis algorithm runs in two steps. \revise{First, we analyze the nested loop of the mapping to split the whole data space into small data spaces until the target level (i.e., bank-level).} The loop analysis runs in an up-down loop order to generate small data spaces by splitting data spaces from the upper loop level. Second, we map the generated data spaces into the correct temporal and spatial locations. The spatial index for each data space can be computed straightforwardly by tracking all spatial loops (i.e., $parallel\_for$).

For the temporal index, we deduce a formula to translate the indices of loop iterations into the temporal steps at each hardware level. Assuming $k$ is the index of the $n^{th}$ temporal loop iteration and $i$ is the global index of $0-(n-1)$ temporal loops, the temporal index of data spaces in the $n^{th}$ loop can be found by:
\begin{equation} \label{eq:2}
  \ S_{k}^{i}(n) = S_{num_{n}}^{i-1}(n)+ G(n) * {k} 
\end{equation}
where $G(n)$ could be found by utilizing Equation~\ref{eq:1}. \revise{We compare them with original data spaces generated from Timeloop~\cite{timeloop} using data space generation with no extrapolating to verify our analytical data spaces.}

\revise{Compared to the strategy in Timeloop's}, which depends on recursive function calls, our method \revise{more efficiently compute all data spaces} in $O(n)$ time complexity, where $n$ is the total number of data spaces. If we implement the data space generation in the Timeloop's recursive function calls, the analysis for one mapping takes around 600 seconds. Our proposed analytical calculation only takes less than 60 seconds.

\subsection{Overlapping Definition}
\label{sec:overlap}
In this section, we clearly define the condition of overlap to prepare for overlap analysis. With the fine-grained data spaces for two consecutive layers, Layer $n$ and $n+1$, we analyze the overlap and estimate the overlapped performance. We denote $O_{t}^{i}$ and $I_{t}^{i}$ as the whole output and input operation spaces in $t^{th}$ temporal step of all hardware instances for Layer $i$. We first find the ready time of $I_{t}^{n+1}$, which is the time when all data in $I_{t}^{n+1}$ are finished by the previous layer (Layer $n$). For each $I_{t_i}^{n+1}$, we need to check all $O_{t}^{n}$ and find the latest time step $t_o$ that $O_{t_o}^{n}$ has an overlap with $I_{t_i}^{n+1}$. If $t_o$ is earlier than the end time of Layer $n$, we can compute $O_{t_i}^{n+1}$ right after $t_o$ because the whole $I_{t_i}^{n+1}$ has been calculated. The computation of $O_{t_i}^{n+1}$ is overlapped with computations in Layer $n$ after $t_o$.

Then, with available instances, the process starts earlier with partial input to exploit overlapping execution between two layers. The ready timestamps $T_{t}$ for input data spaces at $t_{n+1}$ are determined by finding the latest output data space of Layer $n$ which intersects (unions) with any input data spaces of Layer $n+1$ for any instances at time step $t_{n+1}$. \revise{Our new evaluation considers the overlapped data spaces (computations) based on the hardware constraints to determine the overlapped performance as the new optimization metric.}

\subsection{Overlap Analysis with Analytical Algorithm} 
\label{sec:fast}

The extensive parallelism exploited through computational overlap gains great performance improvement. To benefit from this advantage, \Design develops the overlap analysis algorithm to determine the accurate overlap time step results for evaluating the \revise{overlapped performance}. In the PIM architecture, we conduct the overlap analysis at the bank level because the analysis at an upper (e.g., channel) produces too coarse-grained data spaces while a lower level (e.g., column) has too many instances that will make the analysis intractable.

\revise{Our previous work, OverlaPIM~\cite{overlapim}, utilizes the naive approach to determine the computational overlap by traversing through all data spaces at a target storage level and finding the ready timestamps.} However, this analysis requires $Q(N*M)$ time complexity with overheads, where $N$ and $M$ represent the number of data spaces for Layer $k$ and Layer $k+1$ separately. Given the massive number of data spaces in PIM DNN acceleration, over \revise{$10^{7}$} in Bank level for some cases, this approach becomes unacceptably expensive and will strongly limit the search space for the framework due to the unsupportable large computation requirement during the runtime.

\revise{To} develop an applicable framework, we proposed a fast analytical computational overlap analysis algorithm to implement \Design. First, as we mentioned in Section~\ref{sec:space}, we utilize Equation~\ref{eq:1} to calculate the index incremental size for each loop level for the Layer $N$, in order words, the preceding layer. Then, for conducting the overlap analysis at the bank level, we traverse through the loops of Layer $N$ in an up-down order from DRAM to Bank level to get the finishing time step for the input data spaces of Layer $N+1$: $$Input^{m} = [K_{min}^{m}, P_{min}^{m}, Q_{min}^{m}: K_{max}^{m}, P_{max}^{m}, Q_{max}^{m}]$$
which is the output from Layer $N$.

During the traversal, we first propose an equation to determine the relationship between the loop and the increment of the lower bound for the data spaces:
\begin{equation} \label{eq:3} 
D(k, p, q)^{n} =
\begin{cases}
     D(k/num_{n} ,p, q)^{n+1}, & \text{if } L(n) \subseteq K, \\
     D(k ,p/num_{n}, q)^{n+1}, & \text{if } L(n) \subseteq P, \\
     D(k ,p, q/num_{n})^{n+1}, & \text{if } L(n) \subseteq Q.
\end{cases}
\end{equation}
where $n$ represents the $n^{th}$ in a down-up order loop, $D(k, p, q)^{n}$ is a list that stores the size for each $K$, $P$, $Q$ tensor element, and we denote $L(n) \subseteq K/P/Q$ when $L(n)$ decomposes the corresponding tensor.

Since each loop decomposes one dimension among $P$, $Q$, and $K$, in the following discussion, for simplicity, for each loop level, we utilize a single variable $d$ to represent the decomposed $K_{min}^{m}, P_{min}^{m}, Q_{min}^{m}$ dimension and ignore the rest.
We define the $S(d)$ to store the lower bound of the decomposed data space and assign the initial value to $d$:
\begin{equation} \label{eq:4}
    \ S(d) = d
\end{equation}
which will be updated during the up-down loop traverse.

We apply the following functions on each loop level $i$ to find the time step and the spatial element index of the output/input of Layer $N$/Layer $N+1$.

If $L(i)$ is a spatial loop, then
\begin{equation} \label{eq:5-1} 
\begin{cases}
    S+=(d-S(d))//D(d),  \\
    S(d) += (d-S(d))//D(d) * D(d)
\end{cases}  
\end{equation}

If $L(i)$ is a temporal loop, then
\begin{equation} \label{eq:5-2}
\begin{cases}
    T+=(d-S(d))//D(d) * G(i) \\
    S(d) += (d-S(d))//D(d) * D(d).
\end{cases}
\end{equation}

where $G(i)$ is found using Equation~\ref{eq:1}, $S$ denotes the lower bound for the spatial index of the input data space (i.e., the ID for the spatial element that generates the corresponding output from Layer $N$), $T$ denotes the lower bound for the temporal index of the input data space, which is the time step we referred to, and $D(x)$ represents the largest data space to awaiting for decomposing. The unmentioned dimensions remain unchanged in the equations. The calculation of $T$ is similar to Equation~\ref{eq:2}. \revise{To take the time steps required for filter multiplication for each decomposed data space into account, we also specifically trace the loop sizes for loop levels that decompose the weights (i.e. $R$ and $S$).} The total sizes will be added to the temporal index for the finalized time step.

\subsection{Overlap-driven Mapping Transformation}
Although computational overlap enhances performance, the overhead for the analysis is non-trivial even with our lightweight algorithms. Therefore, it is critical to find a new method to search for more mappings in each overlap analysis.
If we can generate the analysis statistics for different mappings in trivial time, we can significantly enlarge the search space, hence improving the search result.

\begin{figure}[t!]
    \centering
    \epsfig{file=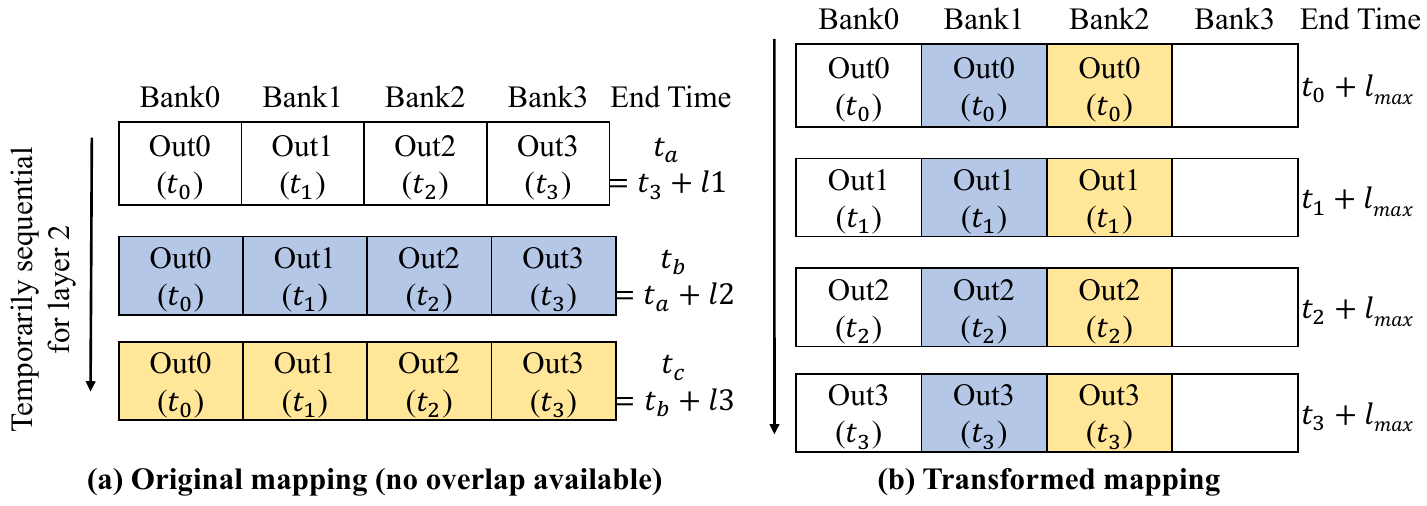, width=\columnwidth}
    \caption{Overlap-based transformation.}
    \label{fig::transform}
\end{figure}

We propose an overlap-based mapping transformation that can directly generate the evaluation results for different mappings based on the mapping that is analyzed in detail. Figure~\ref{fig::transform} shows an example of mapping transformation, where we mark the ready time for the input of each data space ($t_x$). The left part shows the original mapping, where no overlap is available because the latest ready time of all data spaces in a time step is the end time of the previous layer (i.e., $t_3$). For transformation, we reorganize the data spaces and reschedule data spaces at the same ready time, as shown in the right part, significantly decreasing the end time of the layer. 

The transformation runs in two steps. First, it sorts the data spaces in the ascending order of ready time of input. \revise{After sorting, the algorithm allocates the memory resources for each data space based on the ready time. In practice, the number of data spaces with the same ready time can be larger than the number of memory resources. Therefore, a round-robin manner is used to schedule the data space with the same ready time to the same memory location. We should note that the transformation is not overhead-free because it might change the locations of partial sums that require data movements for reduction.} Since the transformation does not require re-analysis of the data space and the complexity of the algorithm is $O(logN)$ bounded by the search, the transformation only introduces \revise{trivial runtime overhead during the process.}

\subsection{Overlap Optimization for the Whole DNN}

\Design supports the mapping search for all layers in the entire DNN model. To evaluate the whole network, we require descriptions of all layers and their corresponding mapping constraints and architecture configurations as inputs. These inputs can be generated automatically through our self-designed toolkit by providing information on architecture design and constraints. The workload parameters for each layer will be auto-generated and processed by \Design following the sequential order for analysis.

During the execution time, the optimization mapper generates candidate mappings based on the configurations, including both the architecture configuration and the mapping constraints. For each mapping, it generates the fine-grained temporal and spatial data spaces, as well as the PIM performance evaluation. Then, the framework calculates the overlap of consecutive layers based on their fine-grained data spaces and recalculates the performance considering the overlapped computations. Based on the overlapped result (i.e. the time step for ready output), \Design also transforms the current mapping into overlap-friendly mappings to increase the search space and improve the performance on latency reduction. The framework continues to update the best mapping based on the overlap-based performance until meeting the termination requirements (similar to Timeloop~\cite{timeloop}), which is set to a fixed number of valid mappings in our framework.

Given that overlapping performance depends on both Layer $n$ and Layer $n+1$, finding the best performance mappings for all layers through searching and comparing mappings would be prohibitively expensive. For example, if we search for $k$ mappings in each layer, the total possible combination of mappings for all $N$ layers would be $k^N$. The exhaustive search for optimal mappings would be unacceptably expensive in this case. Thus, for each Layer $n+1$, the overlapping optimization search is done based on the best mapping found for Layer $n$. Our evaluation shows that such a linear method can produce high-performance mappings. We leave a more in-depth investigation of global optimization for future work. 

We \revise{notice} that the skip connections in ResNet models can also affect the whole network mapping optimization. However, the skip layers can be executed in parallel with other layers ($\geq 2$) from the same block, with careful mapping optimization, the skip layer will be completed during the execution time of other layers. Thus, its mapping will not affect the total latency optimization for the whole DNN network.

\subsection{Search Algorithm Optimization}
\label{sec:search}
\revise{To} better diminish the search space limitation for the whole DNN optimization and provide a more balanced priority for each layer in the optimization process, as mentioned in Section~\ref{sec:flow}, we proposed our novel search strategies. Conventional whole network mapping optimization always analyzes layers starting from the first layer and forwarding to the next following the temporal order\cite{overlapim, pim-dl, pimdnn2}. \revise{However, considering the interdependent relationship for performance between Layer $N$ and Layer $N+1$, searching from the beginning tends to optimize earlier layers better than later layers. With the restricted number of mappings being searched, searching mappings for Layer $N+1$ based on fixed mapping of Layer $N$ creates biased results. To address this, we introduced an interface that enables mapping exploration based on the fixed mapping of Layer $N + 1$ for Layer $N$. This approach offers more mapping search flexibility and reduces timely bias during whole DNN mapping exploration.}

\revise{More specifically, all succeeding layers are restricted to determine the overlap mapping based on the fixed mapping from the proceeding layer to reduce the search workloads from $k^N$ to $N\times k$. However, there exists the possibility that the succeeding layers can generate better overlap mappings based on the sub-optimal mappings of the proceeding layers and provide better throughput among all layers. The cycle performances of NN layers are dominated by data point accesses and data transfer, which will scale with the total number of parameters for each layer. Layers with larger output height and weight ($P$ and $Q$) and a larger number of input and output channels ($C$ and $K$) are more likely to become the performance bottleneck of the whole network optimization. We break the convention of starting generating the optimal mapping for Layer $1$. Our novel search interface allows users to select a middle layer as the starting point for the whole network optimization based on the size of input, output, and weight parameters. Thus, \Design can provide more insights into the whole network optimization within a linear time runtime increment.}

\revise{In conclusion, we proposed 3 search methods to enlarge our search space.} The \revise{``Forward''} method is the conventional method that searches starting from the first layer following the temporal order. The \revise{``Backward''} method starts mapping exploration from the last layer and finds mappings for previous layers based on the fixed mappings for its succeeding layer following the reverse temporal order. \revise{The ``Middle'' method triggers the whole network mapping search beginning at an intermediate layer. For our \Design framework, we explore the ``Middle'' method based on the layer with the largest output size (i.e. $P\times Q \times K$) and with the largest overall size (i.e. $P \times Q \times C \times K$). The ``Forward'' and ``Backward'' searches are conducted separately from the chosen layer.}

\section{Experiments}
\subsection{Experimental Setup}
\subsubsection{Implementation}
\revise{We implemented our proposed framework using Timeloop~\cite{timeloop} with extensive customization}.

\subsubsection{Baseline}
\revise{We compare \Design to several baselines based on state-of-the-art PIM mapping optimization~\cite{pim-dl}.} Specifically, \revise{``Best Original''} indicates the mapping optimized by the original framework and does not consider overlap;  ``Best Original Overlap'' means the same mapping as the ``Best Original'' but the performance considers the overlapped computation, which is analyzed by \Design; ``Best Overlap'' is the mapping optimized based on the execution time considering overlap in the search process (no transformation); ``Best Transform'' indicates the optimized mapping generated by considering the transformation during the overlap-based optimization, which is the result generated by \Design. \revise{``Original Transform''} and \revise{``Overlap Transform''} represent the mapping analyzed by the transformation mechanism, given and not given consideration for overlapped computation respectively. Additionally, \revise{``OverlaPIM Original''}, \revise{``OverlaPIM Original Overlap''}, \revise{``OverlaPIM Overlap''}, and \revise{``OverlaPIM Transform''} means ``Best Original'', ``Best Original Overlap'', ``Best Overlap'', and ``Best Transform'' respectively from our previous work OverlaPIM \cite{overlapim}.

\subsubsection{Architecture Configuration}
We use the HBM2~\cite{hbm} with the support of majority-based bit-serial computation~\cite{ali2019memory} as the base technology for the PIM architecture. We allocate a fixed number of HBM channels (8 banks/channel, 32MB bank) for each layer, depending on the network size. The whole system has 4 8GB HBM2 stacks, with a total of 128 channels. We assume all 4 HBM2 stacks are connected through a host machine with a 256GB/s bus. \revise{We extract the timing and internal/external bandwidth of HBM from previous work~\cite{o2017fine}. The detailed latency and energy for memory commands (e.g., activate, precharge, etc.) are shown in Table~\ref{tab:hbm}.} We carefully co-design between architecture constraints and the PIM performance model. We re-construct the number of instances and entries of each hierarchy level to benefit the runtime performance for the search and evaluation process, while we add approximation and adjustment functions in the PIM performance model to ensure reliable latency results.

\begin{table}[t!]
  \centering
  \caption{\textbf{ARCHITECTURAL PARAMETERS FOR \Design}}
  \begin{tabular}{|c|c|}
    \hline
    \textbf{HBM Organization} & \makecell{Channels/die = 32, Banks/channel = 8,\\ Bank = 32MB}  \\ \hline
    \textbf{HBM Timing(ns)} & \makecell{$t_{RC}=45$, $t_{RCD}=16$,  $t_{RAS}=29$, $t_{CL}=16$, \\ $t_{RRD}=2$, $t_{WR}=16$, $t_{CCD_{s}}=2$, $t_{CCD_{L}}=4$}  \\
    \hline
    \textbf{HMB Energy (pJ)} & \makecell{$e_{ACT}=909$, $e_{Pre-GSA}=1.51$, \\ $e_{Post-GSA}=1.17$, $e_{I/O}=0.80$} \\
    \hline
  \end{tabular}
  \label{tab:hbm}
\end{table}

\subsubsection{Workloads and Mapping Constraints}
We evaluate the efficiency of \Design on three popular DNN networks, ResNet-18~\cite{resnet}, VGG-16~\cite{vgg}, and ResNet-50~\cite{resnet}. \revise{We note that \Design is general-purpose to all workloads as mentioned in Section~\ref{sec:config}.}

\begin{figure}[t!]
    \centering
    \epsfig{file=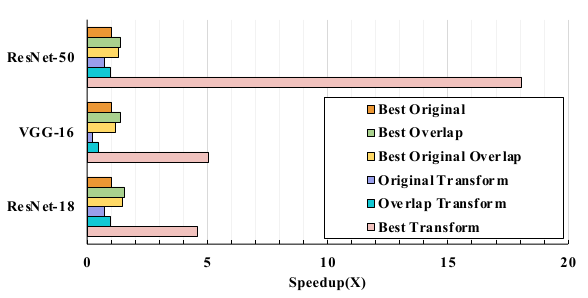, width=\columnwidth}
    \caption{Overall performance comparison over different algorithms.}
    \label{fig:overall_exp}
    \squeezeup
\end{figure}

\begin{figure}[t!]
    \centering
    \epsfig{file=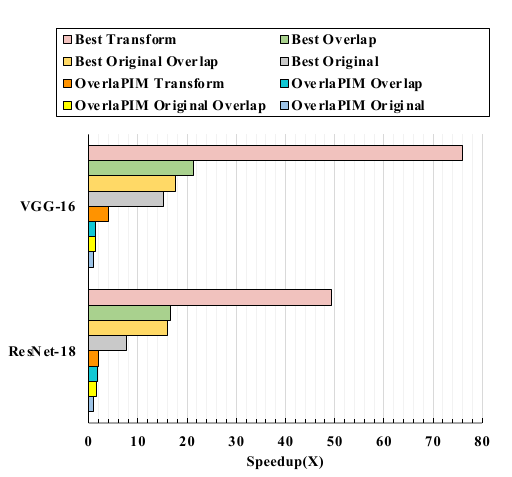, width=\columnwidth}
    \caption{Comparison between OverlaPIM and \Design.}
    \label{fig:overlapim_vs_fast}
    \squeezeup
\end{figure}

\subsection{Overall Comparison}
Figure~\ref{fig:overall_exp} shows the overall results of different mapping optimization algorithms. For ResNet-18, the overlapped latency of the best overlapping mapping (Best Overlap) is $1.6\times$ faster than the end-to-end latency without overlapping (Best Original). If we adopt the transformation in the search process, the best mapping (''Best Transform'') further improves the performance of ``Best Overlap'' by $2.9\times$.  For VGG-16, ``Best Overlap'' provides $1.17\times$ speedup over ``Best Original'', and ''Best Transform'' provides $5.0\times$ speedup over ``Best Original''. On a larger workload, ResNet-50, \Design demonstrates even better performance. \revise{``Best Overlap''} generates $1.3\times$ faster mapping than ``Best Original'' and \revise{``Best Transform''} is $18.1\times$ better than ``Best Original''.

For comparison based on the same mapping with different algorithms, \revise{``Overlap Transform''} and \revise{``Best Transform''} achieve $1.4\times$ and $6.5\times$ speedup over \revise{``Original Transform''} separately in ResNet-18. \Design gives a better improvement in VGG-16 based on this condition. ``Overlap Transform'' provides $2.0\times$ speedup over ``Original Transform''. The overlap-based optimization with transformation (''Best Transform'') produces significantly better mappings, which is $22.0\times$ faster than \revise{``Original Transform''}. $24.5\times$ speedup is gained by \revise{``Best Transform''} over ``Original Transform'' on workload ResNet-50.

\begin{figure*}[t!]
    \centering
    \epsfig{file=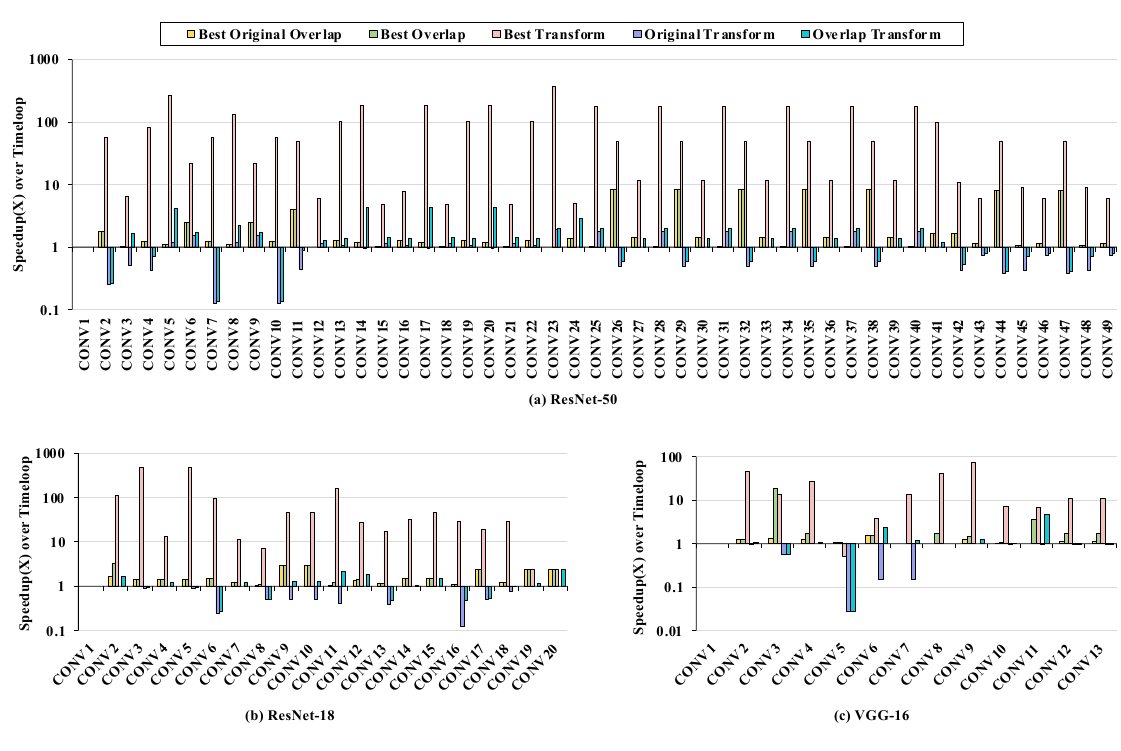, width=\textwidth}
    \caption{The per-layer performance comparison on ResNet-18, VGG-16, and ResNet-50. All results are normalized to ``Best Original''(Timeloop~\cite{timeloop}).}
    \label{fig:per_layer_exp}
\end{figure*}

\subsection{Fast-OverlaPIM versus OverlaPIM}
Figure~\ref{fig:overlapim_vs_fast} displays a performance comparison between our proposed \Design and OverlaPIM \cite{overlapim}. For a fair comparison, we run two tools in the same amount of runtime.
For ResNet-18 and VGG-16, \Design achieves $7.6\times$ and $15.1\times$ better performance on the original cycles (\revise{``Best Original''} over \revise{``OverlaPIM Original''}). This improvement stems from the search space enlargement given the advancement of the runtime performance. With the transformation mechanism, \Design achieves $49.3\times$ to $76.1\times$ speedup over OverlaPIM (\revise{``Best Transform''} over \revise{``OverlaPIM Original''}). The relative speedup of $4.6\times$ and $18.1\times$ for the transform cycle compared to the original cycles from \Design also overcomes previous results of $2.1\times$ to $4.1\times$ performance improvement. The primary factor for this enhancement is that with analytical overlap analysis, \Design supports more detailed and accurate finish time step generation, which benefits transformation. Furthermore, it is important to note that our new framework is now able to handle larger workloads, such as ResNet-50, due to the significantly faster runtime (Section~\ref{sec:runtime}).

\subsection{Per-layer Breakdown}
Figure~\ref{fig:per_layer_exp} shows the per-layer performance comparison over different mapping optimization algorithms. The figures are produced over a logarithm scale as some layers achieve very significant performance benefits. Figure~\ref{fig:per_layer_exp} (a) displays the results on ResNet-50. As we observe, ``Best Overlap'' improves the performance largely (more than $2\times$ speedup) for only 10 out of 49 layers. On the other hand, ``Best Transform'' improves the performance from $4.8\times$ to $369.2\times$.
Figure~\ref{fig:per_layer_exp} (b) is the per-layer performance breakdown for ResNet-18. As shown in the figure, the overlap-based algorithms, both ``Best Overlap'' and ``Best Transform'', find better mappings than the original method (''Best Original''). ``Best Transform'' gives at least $2.3\times$ and at up to $474.3\times$ speedup among Layer $2$ to Layer $20$. The proposed optimization achieves small performance improvements when the existing mapping algorithm (without overlap consideration) coincidentally produces mappings with a high overlap ratio, as shown in Figure~\ref{fig:motivation}. Figure~\ref{fig:per_layer_exp}(c) is the per-layer performance comparison on VGG-16. Even though 4 out of 13 layers fail to find better mappings with \revise{``Best Overlap''} compared to \revise{``Best Original''}. The mapping with transformation still shows a significant benefit of transformation, which is $3.8\times$ to $74.7\times$ speedup over the original method. 

However, we can also observe that for \revise{``Original Transform''} and \revise{``Overlap Transform''}, all 3 workloads generate mapping with worse performance compared to \revise{``Best Original''}. This observation also supports our motivation that the best mapping explored by Timeloop may not necessarily be the best mapping with overlapped computation.

\subsection{Sensitivity Analysis of Memory Capacity}
\label{sensitivity}
\begin{figure}[t!]
    \centering
    \epsfig{file=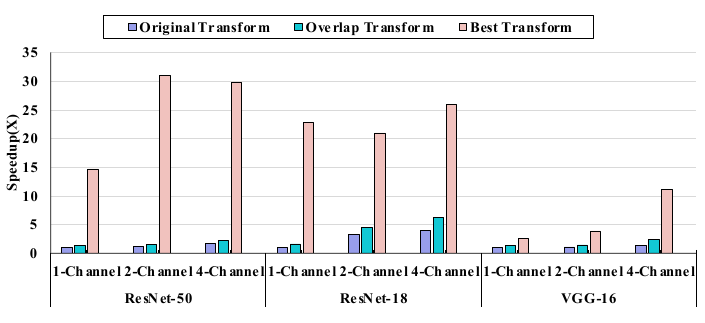, width=\columnwidth}
    \caption{Efficiency of \Design on different memory capacity. All results are normalized to the ``Best Original'' time for the 1-channel setting.}
    \label{fig:arch_sense}
\end{figure}

We adopt \Design on various architecture settings where we allocate different amounts of memory resources for each layer. We compare the performance of different optimization algorithms in various architecture settings, as shown in Figure~\ref{fig:arch_sense}. For the sake of clarity and simplicity, we compared our search methods based on \revise{``Original Transform''}, \revise{``Overlap Transform''}, and \revise{``Best Transform''} as they resulted from the same series of mappings.

The results show that, in ResNet-18, ``Best Transform'' is 6.5 $\times$ and 1.6$\times$ faster than \revise{``Original Transform''}, and \revise{``Overlap Transform''} for the 4-channel setting, similar but slightly better than the results shown in Figure~\ref{fig:overall_exp} (the default 2-channel setting). The performance improvements on the 1-channel setting appear to be beneficial for the ``Best Transform'' algorithm. It achieves 1.6$\times$ and 22.8$\times$ speedup over the baseline. Across different settings, we observe similar performance on VGG-16.
\Design also shows its memory sensitivity on our larger workload, ResNet-50. As shown in Figure~\ref{fig:arch_sense}, the whole network ResNet50 demonstrates the best speedup for \revise{``Overlap Transform''} and ``Best Transform'', where 2-channel and 4-channel settings have similar performance improvement on ``Best Transform'', $31.0 \times$ and $29.9 \times$, separately. The speedup for the 1-channel setting for ``Best Transform'' in ResNet-50 is $14.6 \times$. The performance improvements with \revise{``Overlap Transform''} are $1.3 \times$, $1.5 \times$, and $2.2 \times$ for each architecture setting. Such results prove that \Design is scalable and general for DNN mapping on PIM architectures.

\subsection{Runtime Comparison} \label{sec:runtime}
\begin{figure}[t!]
    \centering
    \epsfig{file=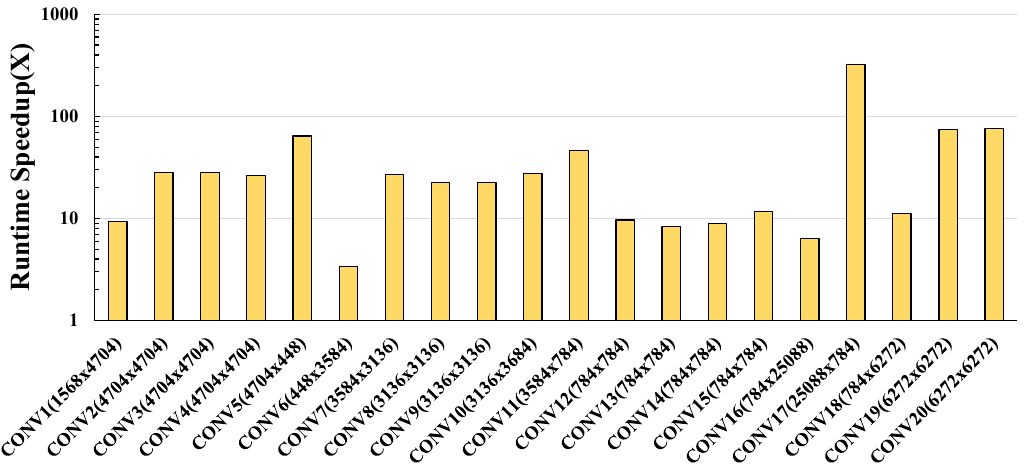, width=\columnwidth}
    \caption{Runtime improvement over OverlaPIM~\cite{overlapim}. AxB in the parentheses means the number of data spaces for Layer $N$ multiplied by the number of data spaces for Layer $N+1$.}
    \label{fig:runtime}
\end{figure}

As mentioned in Section~\ref{sec:fast}, previous work \cite{overlapim} conducts overlap analysis by exhaustively comparing all data spaces of Layer $N$ and Layer $N+1$, which makes the analysis unacceptably expensive. Our Analytical Computational Overlap Analysis greatly improves the runtime efficiency. 
\revise{Figure~\ref{fig:runtime} shows that the algorithm from \Design achieves $3.4\times$ to $323.1\times$ runtime speedup. As the number of data spaces gets larger, the speedup increases more than quadratically.} 

\subsection{Search Method Comparison}
We experimented with our 3 search methods proposed in Section~\ref{sec:search} on ResNet-18, VGG-16, and ResNet-50. Again, similarly to Section~\ref{sensitivity}, for the sake of clarity and simplicity, we compared our search methods based on \revise{``Original Transform''}, \revise{``Overlap Transform''}, and \revise{``Best Transform''}.

\begin{figure}[t!]
    \centering
    \epsfig{file=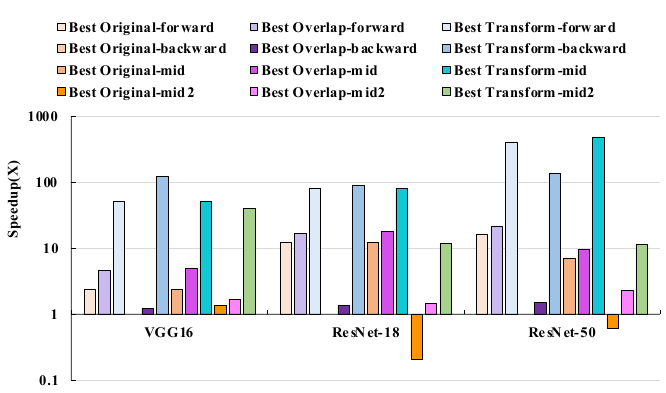, width=\columnwidth}
    \caption{\revise{The overall performance comparison on ResNet-18 and VGG-16, and ResNet-50 on different search methods. All results are normalized to ``Best Original'' with the ``Backward'' method.}}
    \label{fig:search_exp}
\end{figure}

As shown in Figure~\ref{fig:search_exp}, although the \revise{``Backward''} method performs worst without transformation (\revise{``Best Original''} and \revise{``Best Overlap''}), on both ResNet-18 and VGG-16, \revise{``Best Transformation''} with \revise{``Backward''} method gains $1.1\times$ and $2.3\times$ better performance than with \revise{``Forward''} method separately. The performance for different search methods on ResNet-50 is different from ResNet-18 and VGG-16. The \revise{``Middle''} method achieves \revise{up to} $1.2 \times$ better performance than \revise{``Forward''} with transformation, whereas \revise{``Forward''} has $2.9 \times$ better performance compared to \revise{``Backward''}. However, for \revise{``Best Overlap''}, \revise{``Forward''} is $2.2 \times$ better than the \revise{``Middle''}.

\revise{Specifically, we explored the ``Middle'' with two heuristics as mentioned in Section~\ref{sec:search}, the choice of the intermediate layer to start the mapping search process is determined based on the layer with the largest output size (i.e. $P\times Q \times K=mid$) and with the largest overall input/output size i.e. $P \times Q \times C \times K=mid2$). The chosen starting layers are Layer$4$ and Layer$18$, Layer$2$ and Layer$9$, and Layer$10$ and Layer$41$ for ResnNet-18, VGG-16, and ResNet-50 respectively. We observed that the performance between two heuristic reaches is up to $41.9 \times$.}
\revise{These results are} very intriguing, which further supports our claim that the optimal mappings from conventional frameworks may not be the best mappings under PIM architecture with overlapping. 

During our experiments, we also observed an insight that with different search methods, the same layers from the same NN model can have very different performances. For ResNet-18, 16 out of 20 layers generate different mappings with different search methods. And for VGG-16, 12 out of 13 layers produce different mappings. 

\highlight{\subsection{Architectural Applicability for ReRAM}}
\begin{figure}[t!]
    \centering
    \epsfig{file=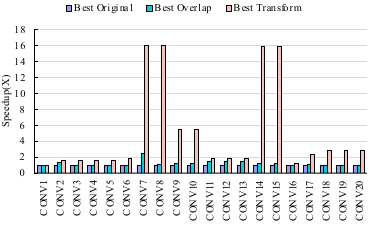, width=\columnwidth}
    \caption{\highlight{The per-layer performance comparison on ResNet-18 with ReRAM.}}
    \label{fig:reram_result}
\end{figure}

\highlight{As mentioned in Section~\ref{sec:config} and Section~\ref{sec:otherpim}, \Design is a versatile PIM-based mapping optimization that can accommodate various PIM-based architectures. Figure~\ref{fig:reram_result} presents the per-layer performance comparison of the ReRAM-based PIM architecture on ResNet-18. ReRAM-based PIM \Design achieves $1.16\times$ overall speedup for ``Best Overlap'' and $2.42\times$ overall speedup for ``Best Transform'', which demonstrates that \Design can support PIM architectures other than DRAM-based PIM.}

\begin{figure}[t!]
    \centering
    \epsfig{file=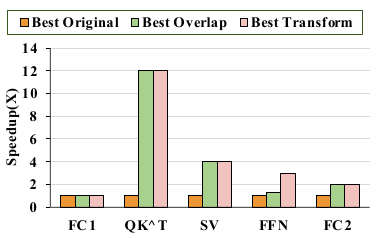, width=0.8\columnwidth}
    \caption{\revise{The per-layer performance comparison on one encoder in BERT encoder.}}
    \label{fig:transformer}
\end{figure}

\revise{\section{Case Study: Self-Attention in Transformers}}
\revise{Utilization of the attention mechanism in Transformers leads to substantial performance improvements in various machine learning tasks. There exists a rising accelerating demand for Transformers\cite{transpim}. Our \Design framework has the potential to support operators such as the self-attention layer (SA). The encoder block in transformers
has three sub-layers including a fully-connected layer (FC),
a self-attention layer, and a feed-forward layer (FFN). The key operation in an SA layer is the scaled dot-product attention computing dependencies between input tokens as Softmax$(\frac {QK^T}{\sqrt{D}})V$. Similar to \cite{timeloop}, by setting R, S, P, and Q to 1, matrix-matrix multiplications can be expressed in \Design; and matrix-vector multiplications can also be expressed by setting R, S, P, Q, and N to 1 for supporting FC and FFN layers, which account for a majority of the computation in transformers. Therefore, \Design supports overlap analysis for the main components of the self-attention layer. Figure~\ref{fig:transformer} illustrates the overlapping results for one encoder layer in BERT \cite{bert}, \Design achieves from $1.3 \times$ to $12.0 \times$ speedup compared to ``Best Orginal''. Since matrix-matrix and matrix-vector multiplication require smaller mapping spaces due to the shallower nested loop than convolution, we notice that our transformation strategy mostly achieves similar performance as overlap without transformation. As the overlap results already exploit parallelism.} 

\section{Conclusion}
\revise{This work proposes a novel DNN mapping framework for PIM, \Design, considering the overlapping computation.} \revise{With customized architecture configuration, \Design provides support as a general PIM NN mapping framework (e.g. DRAM-based, ReRAM-based, SRAM-based).  \Design integrates the overlap analysis in the mapping optimization that searches for the best DNN mapping based on the overlapped latency, instead of sequential latency by the existing methods.} We propose several techniques to improve the effectiveness and efficiency of \Design, including a fine-grained operation space generation, an overlap-based performance analysis, and a transformation algorithm to quickly find overlap-friendly mappings. \Design can find mappings that achieve 4.6$\times$ to 18.1$\times$ better performance than mappings optimized by existing methods.

\section*{Acknowledgment}
This work was funded by PRISM, one of seven centers in JUMP 2.0 (an SRC program sponsored by DARPA), SRC Global Research Collaboration (GRC) grant, and NSF grants \#2112167, \#2003279, \#2100237, \#2112665, and \#2052809.
\bibliographystyle{IEEEtran}
\bibliography{refs}

\begin{thebibliography}{10}
\providecommand{\url}[1]{#1}
\csname url@samestyle\endcsname
\providecommand{\newblock}{\relax}
\providecommand{\bibinfo}[2]{#2}
\providecommand{\BIBentrySTDinterwordspacing}{\spaceskip=0pt\relax}
\providecommand{\BIBentryALTinterwordstretchfactor}{4}
\providecommand{\BIBentryALTinterwordspacing}{\spaceskip=\fontdimen2\font plus
\BIBentryALTinterwordstretchfactor\fontdimen3\font minus \fontdimen4\font\relax}
\providecommand{\BIBforeignlanguage}[2]{{%
\expandafter\ifx\csname l@#1\endcsname\relax
\typeout{** WARNING: IEEEtran.bst: No hyphenation pattern has been}%
\typeout{** loaded for the language `#1'. Using the pattern for}%
\typeout{** the default language instead.}%
\else
\language=\csname l@#1\endcsname
\fi
#2}}
\providecommand{\BIBdecl}{\relax}
\BIBdecl

\bibitem{touchgo}
F.~Yang, C.~Ma, J.~Zhang, J.~Zhu, W.~Yuan, and A.~Owens, ``Touch and go: Learning from human-collected vision and touch,'' \emph{arXiv preprint arXiv:2211.12498}, 2022.

\bibitem{videohuman}
D.~Wu, N.~Sharma, and M.~Blumenstein, ``Recent advances in video-based human action recognition using deep learning: A review,'' in \emph{2017 International Joint Conference on Neural Networks (IJCNN)}.\hskip 1em plus 0.5em minus 0.4em\relax IEEE, 2017, pp. 2865--2872.

\bibitem{image1}
H.~Zhao, F.~Yang, X.~Fu, and X.~Li, ``Rbc: Rectifying the biased context in continual semantic segmentation,'' in \emph{Computer Vision--ECCV 2022: 17th European Conference, Tel Aviv, Israel, October 23--27, 2022, Proceedings, Part XXXIV}.\hskip 1em plus 0.5em minus 0.4em\relax Springer, 2022, pp. 55--72.

\bibitem{iot1}
X.~Wang, X.~Wang, and S.~Mao, ``Rf sensing in the internet of things: A general deep learning framework,'' \emph{IEEE Communications Magazine}, vol.~56, no.~9, pp. 62--67, 2018.

\bibitem{iot2}
M.~A. Al-Garadi, A.~Mohamed, A.~K. Al-Ali, X.~Du, I.~Ali, and M.~Guizani, ``A survey of machine and deep learning methods for internet of things (iot) security,'' \emph{IEEE Communications Surveys \& Tutorials}, vol.~22, no.~3, pp. 1646--1685, 2020.

\bibitem{iot3}
X.~Yu, R.~Yu, J.~Yang, and X.~Duan, ``A robotic auto-focus system based on deep reinforcement learning,'' in \emph{2018 15th International Conference on Control, Automation, Robotics and Vision (ICARCV)}, 2018, pp. 204--209.

\bibitem{nlp1}
D.~W. Otter, J.~R. Medina, and J.~K. Kalita, ``A survey of the usages of deep learning for natural language processing,'' \emph{IEEE transactions on neural networks and learning systems}, vol.~32, no.~2, pp. 604--624, 2020.

\bibitem{nlp2}
X.~Liu, P.~He, W.~Chen, and J.~Gao, ``Multi-task deep neural networks for natural language understanding,'' \emph{arXiv preprint arXiv:1901.11504}, 2019.

\bibitem{resnet}
K.~He, X.~Zhang, S.~Ren, and J.~Sun, ``Deep residual learning for image recognition,'' in \emph{2016 IEEE Conference on Computer Vision and Pattern Recognition (CVPR)}, 2016, pp. 770--778.

\bibitem{floatpim}
M.~Imani, S.~Gupta, Y.~Kim, and T.~Rosing, ``Floatpim: In-memory acceleration of deep neural network training with high precision,'' in \emph{2019 ACM/IEEE 46th Annual International Symposium on Computer Architecture (ISCA)}, 2019, pp. 802--815.

\bibitem{gram}
M.~Zhou, M.~Imani, S.~Gupta, Y.~Kim, and T.~Rosing, ``Gram: graph processing in a reram-based computational memory,'' in \emph{IEEE Asia and South Pacific Design Automation Conference}, 2019.

\bibitem{transpim}
M.~Zhou, W.~Xu, J.~Kang, and T.~Rosing, ``Transpim: A memory-based acceleration via software-hardware co-design for transformer,'' in \emph{2022 IEEE International Symposium on High-Performance Computer Architecture (HPCA)}, 2022, pp. 1071--1085.

\bibitem{felix}
T.~Soliman, N.~Laleni, T.~Kirchner, F.~M{\"u}ller, A.~Shrivastava, T.~K{\"a}mpfe, A.~Guntoro, and N.~Wehn, ``Felix: A ferroelectric fet based low power mixed-signal in-memory architecture for dnn acceleration,'' \emph{ACM Transactions on Embedded Computing Systems}, vol.~21, no.~6, pp. 1--25, 2022.

\bibitem{pimdnn1}
L.~Yang, Z.~He, S.~Angizi, and D.~Fan, ``Processing-in-memory accelerator for dynamic neural network with run-time tuning of accuracy, power and latency,'' in \emph{2020 IEEE 33rd International System-on-Chip Conference (SOCC)}, 2020, pp. 117--122.

\bibitem{pimdnn2}
S.~Angizi, Z.~He, D.~Reis, X.~S. Hu, W.~Tsai, S.~J. Lin, and D.~Fan, ``Accelerating deep neural networks in processing-in-memory platforms: Analog or digital approach?'' in \emph{2019 IEEE Computer Society Annual Symposium on VLSI (ISVLSI)}, 2019, pp. 197--202.

\bibitem{timeloop}
\BIBentryALTinterwordspacing
A.~Parashar, P.~Raina, Y.~S. Shao, Y.-H. Chen, V.~A. Ying, A.~Mukkara, R.~Venkatesan, B.~Khailany, S.~W. Keckler, and J.~Emer, ``Timeloop: A systematic approach to dnn accelerator evaluation,'' in \emph{2019 IEEE international symposium on performance analysis of systems and software (ISPASS)}.\hskip 1em plus 0.5em minus 0.4em\relax IEEE, 2019, pp. 304--315. [Online]. Available: \url{https://github.com/NVlabs/timeloop}
\BIBentrySTDinterwordspacing

\bibitem{pim-dl}
M.~Zhou, G.~Chen, M.~Imani, S.~Gupta, W.~Zhang, and T.~Rosing, ``Pim-dl: Boosting dnn inference on digital processing in-memory architectures via data layout optimizations,'' in \emph{2021 30th International Conference on Parallel Architectures and Compilation Techniques (PACT)}.\hskip 1em plus 0.5em minus 0.4em\relax IEEE, 2021, pp. 1--1.

\bibitem{sparseloop}
Y.~N. Wu, P.-A. Tsai, A.~Parashar, V.~Sze, and J.~S. Emer, ``Sparseloop: An analytical approach to sparse tensor accelerator modeling,'' 2023.

\bibitem{eyeriss}
Y.-H. Chen \emph{et~al.}, ``Eyeriss: A spatial architecture for energy-efficient dataflow for convolutional neural networks,'' \emph{ACM SIGARCH computer architecture news}, 2016.

\bibitem{MAESTRO}
H.~Kwon, P.~Chatarasi, M.~Pellauer, A.~Parashar, V.~Sarkar, and T.~Krishna, ``Understanding reuse, performance, and hardware cost of dnn dataflow: A data-centric approach,'' in \emph{Proceedings of the 52nd Annual IEEE/ACM International Symposium on Microarchitecture}, 2019, pp. 754--768.

\bibitem{interstellar}
X.~Yang, M.~Gao, Q.~Liu, J.~Setter, J.~Pu, A.~Nayak, S.~Bell, K.~Cao, H.~Ha, P.~Raina \emph{et~al.}, ``Interstellar: Using halide's scheduling language to analyze dnn accelerators,'' in \emph{Proceedings of the Twenty-Fifth International Conference on Architectural Support for Programming Languages and Operating Systems}, 2020, pp. 369--383.

\bibitem{ruby}
M.~Horeni, P.~Taheri, P.-A. Tsai, A.~Parashar, J.~Emer, and S.~Joshi, ``Ruby: Improving hardware efficiency for tensor algebra accelerators through imperfect factorization,'' in \emph{2022 IEEE International Symposium on Performance Analysis of Systems and Software (ISPASS)}, 2022, pp. 254--266.

\bibitem{huang2021cosa}
Q.~Huang, M.~Kang, G.~Dinh, T.~Norell, A.~Kalaiah, J.~Demmel, J.~Wawrzynek, and Y.~S. Shao, ``Cosa: Scheduling by constrained optimization for spatial accelerators,'' in \emph{2021 ACM/IEEE 48th Annual International Symposium on Computer Architecture (ISCA)}.\hskip 1em plus 0.5em minus 0.4em\relax IEEE, 2021, pp. 554--566.

\bibitem{anadnn}
Y.~Zhao, C.~Li, Y.~Wang, P.~Xu, Y.~Zhang, and Y.~Lin, ``Dnn-chip predictor: An analytical performance predictor for dnn accelerators with various dataflows and hardware architectures,'' in \emph{ICASSP 2020-2020 IEEE International Conference on Acoustics, Speech and Signal Processing (ICASSP)}.\hskip 1em plus 0.5em minus 0.4em\relax IEEE, 2020, pp. 1593--1597.

\bibitem{puma}
A.~Ankit, I.~E. Hajj, S.~R. Chalamalasetti, G.~Ndu, M.~Foltin, R.~S. Williams, P.~Faraboschi, W.-m.~W. Hwu, J.~P. Strachan, K.~Roy \emph{et~al.}, ``Puma: A programmable ultra-efficient memristor-based accelerator for machine learning inference,'' in \emph{Proceedings of the Twenty-Fourth International Conference on Architectural Support for Programming Languages and Operating Systems}, 2019, pp. 715--731.

\bibitem{prime}
P.~Chi, S.~Li, C.~Xu, T.~Zhang, J.~Zhao, Y.~Liu, Y.~Wang, and Y.~Xie, ``Prime: A novel processing-in-memory architecture for neural network computation in reram-based main memory,'' \emph{ACM SIGARCH Computer Architecture News}, vol.~44, no.~3, pp. 27--39, 2016.

\bibitem{yang_design}
T.-J. Yang and V.~Sze, ``Design considerations for efficient deep neural networks on processing-in-memory accelerators,'' in \emph{2019 IEEE International Electron Devices Meeting (IEDM)}.\hskip 1em plus 0.5em minus 0.4em\relax IEEE, 2019, pp. 22--1.

\bibitem{z-pim}
J.-H. Kim, J.~Lee, J.~Lee, J.~Heo, and J.-Y. Kim, ``Z-pim: A sparsity-aware processing-in-memory architecture with fully variable weight bit-precision for energy-efficient deep neural networks,'' \emph{IEEE Journal of Solid-State Circuits}, vol.~56, no.~4, pp. 1093--1104, 2021.

\bibitem{overlapim}
M.~Zhou, X.~Wang, and T.~Rosing, ``Overlapim: Overlap optimization for processing in-memory neural network acceleration,'' in \emph{2023 Design, Automation \& Test in Europe Conference \& Exhibition (DATE)}, 2023, pp. 1--6.

\bibitem{vgg}
K.~Simonyan and A.~Zisserman, ``Very deep convolutional networks for large-scale image recognition,'' \emph{arXiv preprint arXiv:1409.1556}, 2014.

\bibitem{wang2019computesram}
J.~Wang, X.~Wang, C.~Eckert, A.~Subramaniyan, R.~Das, D.~Blaauw, and D.~Sylvester, ``A 28-nm compute sram with bit-serial logic/arithmetic operations for programmable in-memory vector computing,'' \emph{IEEE Journal of Solid-State Circuits}, vol.~55, no.~1, pp. 76--86, 2019.

\bibitem{drisa}
S.~Li, D.~Niu, K.~T. Malladi, H.~Zheng, B.~Brennan, and Y.~Xie, ``Drisa: A dram-based reconfigurable in-situ accelerator,'' in \emph{Annual IEEE/ACM International Symposium on Microarchitecture (MICRO)}, 2017, pp. 288--301.

\bibitem{simdram}
N.~Hajinazar, G.~F. Oliveira, S.~Gregorio, J.~Ferreira, N.~M. Ghiasi, M.~Patel, M.~Alser, S.~Ghose, J.~G. Luna, and O.~Mutlu, ``Simdram: An end-to-end framework for bit-serial simd computing in dram,'' 2021.

\bibitem{gao2019computedram}
F.~Gao, G.~Tziantzioulis, and D.~Wentzlaff, ``Computedram: In-memory compute using off-the-shelf drams,'' in \emph{Proceedings of the 52nd annual IEEE/ACM international symposium on microarchitecture}, 2019, pp. 100--113.

\bibitem{ali2019memory}
M.~F. Ali, A.~Jaiswal, and K.~Roy, ``In-memory low-cost bit-serial addition using commodity dram technology,'' \emph{IEEE Transactions on Circuits and Systems I: Regular Papers}, vol.~67, no.~1, pp. 155--165, 2019.

\bibitem{dpsim}
M.~Zhou, M.~Imani, Y.~Kim, S.~Gupta, and T.~Rosing, ``Dp-sim: A full-stack simulation infrastructure for digital processing in-memory architectures,'' in \emph{Proceedings of the 26th Asia and South Pacific Design Automation Conference}, 2021, pp. 639--644.

\bibitem{reram1}
B.~C. Jang, Y.~Nam, B.~J. Koo, J.~Choi, S.~G. Im, S.-H.~K. Park, and S.-Y. Choi, ``Memristive logic-in-memory integrated circuits for energy-efficient flexible electronics,'' \emph{Advanced Functional Materials}, vol.~28, no.~2, p. 1704725, 2018.

\bibitem{spin-electronics}
S.~Shreya, A.~Jain, and B.~K. Kaushik, ``Computing-in-memory architecture using energy-efficient multilevel voltage-controlled spin-orbit torque device,'' \emph{IEEE Transactions on Electron Devices}, vol.~67, no.~5, pp. 1972--1979, 2020.

\bibitem{ambit}
V.~Seshadri, D.~Lee, T.~Mullins, H.~Hassan, A.~Boroumand, J.~Kim, M.~A. Kozuch, O.~Mutlu, P.~B. Gibbons, and T.~C. Mowry, ``Ambit: In-memory accelerator for bulk bitwise operations using commodity dram technology,'' in \emph{Proceedings of the 50th Annual IEEE/ACM International Symposium on Microarchitecture}, 2017, pp. 273--287.

\bibitem{ambit2}
\BIBentryALTinterwordspacing
V.~Seshadri and O.~Mutlu, ``In-dram bulk bitwise execution engine,'' \emph{ArXiv}, vol. abs/1905.09822, 2019. [Online]. Available: \url{https://api.semanticscholar.org/CorpusID:165163655}
\BIBentrySTDinterwordspacing

\bibitem{hbm}
Y.-C. Kwon, S.~H. Lee, J.~Lee, S.-H. Kwon, J.~M. Ryu, J.-P. Son, O.~Seongil, H.-S. Yu, H.~Lee, S.~Y. Kim \emph{et~al.}, ``A 20nm 6gb function-in-memory dram, based on hbm2 with a 1.2 tflops programmable computing unit using bank-level parallelism, for machine learning applications,'' in \emph{IEEE International Solid-State Circuits Conference (ISSCC)}, vol.~64, 2021, pp. 350--352.

\bibitem{analog}
B.~Taylor, Q.~Zheng, Z.~Li, S.~Li, and Y.~Chen, ``Processing-in-memory technology for machine learning: From basic to asic,'' \emph{IEEE Transactions on Circuits and Systems II: Express Briefs}, vol.~69, no.~6, pp. 2598--2603, 2022.

\bibitem{mnsim}
Z.~Zhu, H.~Sun, T.~Xie, Y.~Zhu, G.~Dai, L.~Xia, D.~Niu, X.~Chen, X.~S. Hu, Y.~Cao \emph{et~al.}, ``Mnsim 2.0: A behavior-level modeling tool for processing-in-memory architectures,'' \emph{IEEE Transactions on Computer-Aided Design of Integrated Circuits and Systems}, 2023.

\bibitem{o2017fine}
M.~O'Connor, N.~Chatterjee, D.~Lee, J.~Wilson, A.~Agrawal, S.~W. Keckler, and W.~J. Dally, ``Fine-grained dram: Energy-efficient dram for extreme bandwidth systems,'' in \emph{Proceedings of the 50th Annual IEEE/ACM International Symposium on Microarchitecture}, 2017, pp. 41--54.

\bibitem{bert}
J.~Devlin, M.-W. Chang, K.~Lee, and K.~Toutanova, ``Bert: Pre-training of deep bidirectional transformers for language understanding,'' \emph{arXiv preprint arXiv:1810.04805}, 2018.

\end{thebibliography}

\section{Biography Section}
\vskip -4\baselineskip plus -1fil
\begin{IEEEbiography}
[{\includegraphics[width=1in,height=1.25in,clip,keepaspectratio]{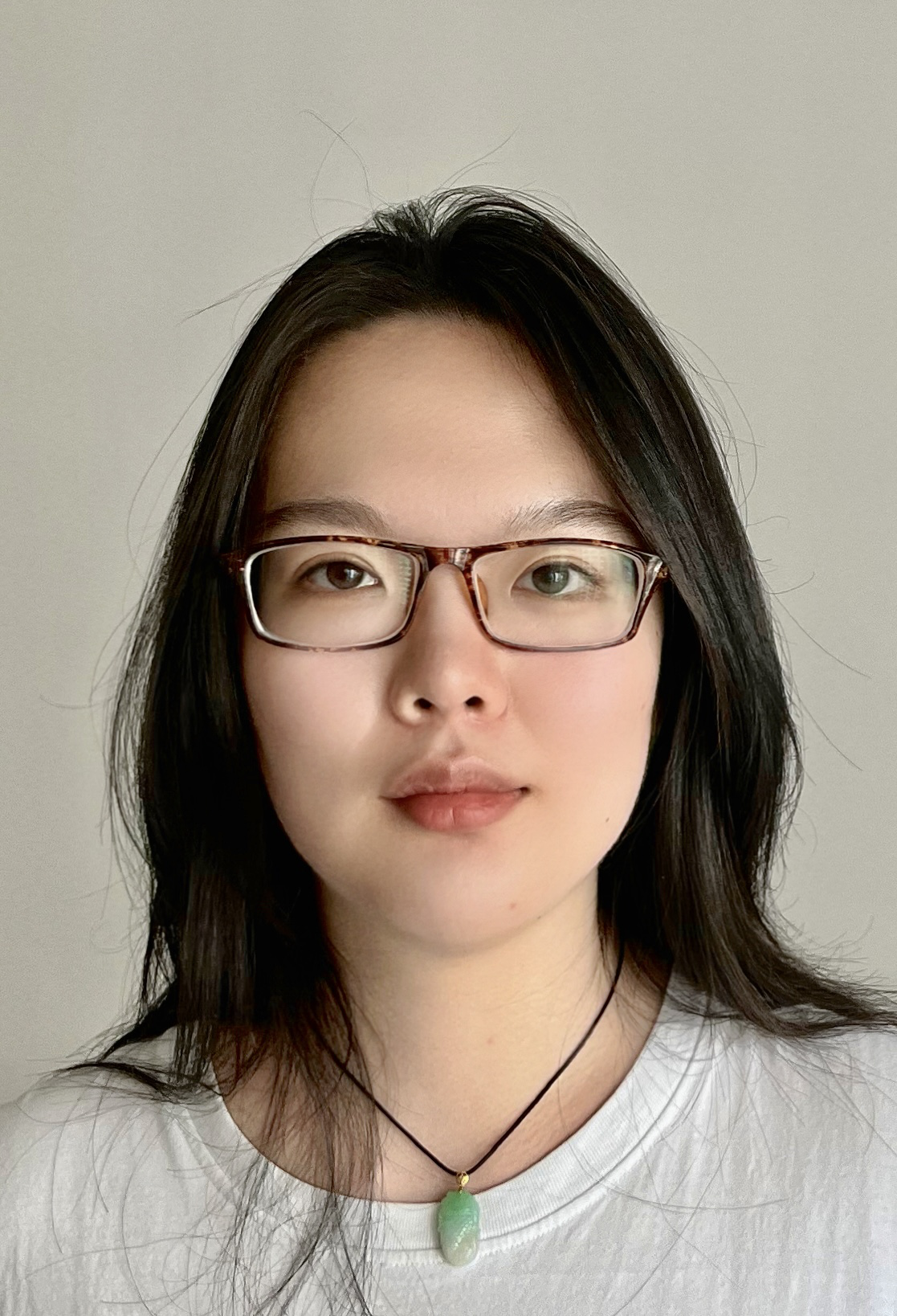}}]{Xuan Wang} 
received the BS degree in Computer Engineering and Mathematics(Applied) from the University of California San Diego, San Diego, CA, USA, in 2019. Currently, she is a first-year Ph.D. Student in the Department of Computer Science and Engineering at the University of California, San Diego. Her research interests include domain-specific acceleration, computer architecture, fully homomorphic encryption, and bioinformatics.
\end{IEEEbiography}
\vskip -3\baselineskip plus -1fil
\begin{IEEEbiography}[{\includegraphics[width=1in,height=1.25in,clip,keepaspectratio]{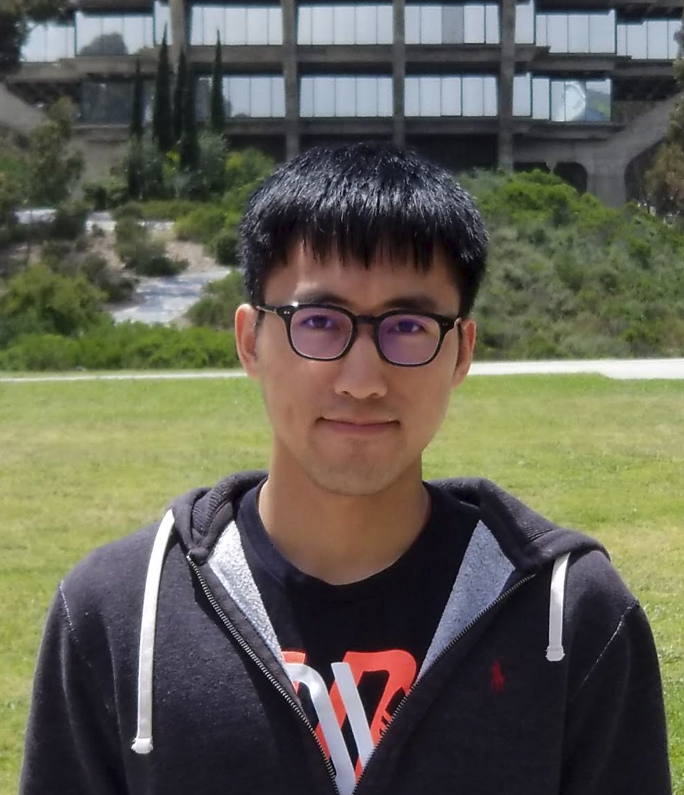}}]{Minxuan Zhou}
received the MS and the PhD degrees in Computer Science from the University of California San Diego, San Diego, CA, USA, in 2017 and 2023, respectively. Currently, he is a Postdoctoral researcher in the Department of Computer Science and Engineering at the University of California, San Diego. His research interests include processing in-memory architecture, system modeling and management, graph processing, and memory systems.
\end{IEEEbiography}
\vskip -3\baselineskip plus -1fil
\begin{IEEEbiography}[{\includegraphics[width=1in,height=1.25in,clip,keepaspectratio]{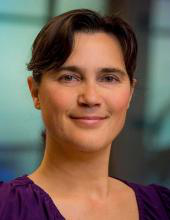}}]{Tajana Rosing}
(Fellow, IEEE) received the MS degree in engineering
management concurrently and the PhD degree from Stanford University, Stanford, CA, USA, in 2001. She is a professor, a holder of the
Fratamico Endowed chair, and the director of System Energy Efficiency Laboratory, University of California at San Diego, La Jolla, CA.
From 1998 to 2005, she was a full-time research scientist with HP
Labs, Palo Alto, CA, while also leading research efforts with Stanford
University, Stanford. She was a senior design engineer with Altera
Corporation, San Jose, CA. She is leading a number of projects,
including efforts funded by DARPA/SRC JUMP CRISP PRISM program with
focus on design of accelerators for analysis of Big Data, DARPA and
NSF funded projects on hyperdimensional computing, and SRC
funded project on IoT system reliability and maintainability. Her current
research interests include energy-efficient computing, cyber–physical,
and distributed systems.
\end{IEEEbiography}
\vskip -3\baselineskip plus -1fil

\end{document}